\documentclass{article}
\usepackage{spconf,amsmath,graphicx,hyperref}
\usepackage{xcolor}
\usepackage{booktabs} 
\usepackage{makecell}
\usepackage{multirow}
\usepackage{amssymb}
\usepackage{cleveref}

\title{Rare Word Recognition and Translation Without Fine-Tuning via Task Vector in Speech Models}
\name{
  Ruihao Jing\sthanks{The work is done with TeleAI.},
  Cheng Gong,
  Yu Jiang,
  Boyu Zhu,
  Shansong Liu,
  Chi Zhang,
  Xiao-Lei Zhang$^{\dagger}$, 
  Xuelong Li\sthanks{Corresponding authors.} 
}

\address{
  Institute of Artificial Intelligence (TeleAI), China Telecom \\
}

\begin{document}
\ninept
\maketitle

\begin{abstract}
Rare words remain a critical bottleneck for speech-to-text systems. While direct fine-tuning improves recognition of target words, it often incurs high cost, catastrophic forgetting, and limited scalability. To address these challenges, we propose a training-free paradigm based on task vectors for rare word recognition and translation. By defining task vectors as parameter differences and introducing word-level task vector arithmetic, our approach enables flexible composition of rare-word capabilities, greatly enhancing scalability and reusability. Extensive experiments across multiple domains show that the proposed method matches or surpasses fine-tuned models on target words, improves general performance by about 5 BLEU, and mitigates catastrophic forgetting.
\end{abstract}
\section{Introduction}
\label{sec:intro}

Large-scale speech models such as Whisper \cite{radford2023robust}, SenseVoice \cite{an2024funaudiollm}, and SeamlessM4T \cite{barrault2023seamlessm4t} have achieved impressive performance by leveraging massive training corpora.
However, the distribution of words in natural language typically exhibits a long-tail pattern, where rare words—such as geographical locations, personal names, and domain-specific terminology—are sparsely covered.
As a result, despite their strong performance in everyday conversational scenarios, these models often struggle to accurately recognize infrequent or specialized vocabulary.
This limitation poses a significant challenge for real-world applications such as automatic speech recognition (ASR) and automatic speech translation (AST), especially in specialized domains such as medicine and law, where rare terminology plays a critical role.

To improve the recognition of rare words in pre-trained speech models, a straightforward approach is to fine-tune the model on rare word datasets.
For example, synthetic speech generated by text-to-speech (TTS) systems can be employed to augment rare-word corpora and subsequently fine-tune pre-trained models \cite{10317116}.
Beyond directly fine-tuning the base model, alternative strategies include fine-tuning auxiliary detection modules or adopting retrieval-augmented generation (RAG) approaches.
CB-Whisper \cite{li2024cb} introduces an additional keyword spotting (KWS) module to identify rare words and feeds them as prompts into the decoding process, improving final transcription accuracy. Similarly, the paper \cite{gaido2023named} leverages a detector to identify domain-specific terms in speech and injects them into the subsequent decoding process. 
In addition, some studies fine-tune a RAG module to estimate the similarity between the current word and rare words stored in an external knowledge base. 
For instance, the translation of rare words can be improved by retrieving translation pairs and leveraging in-context learning \cite{li2024optimizing}. 
Likewise, the method described in \cite{wu2025locate} uses a “locate-and-focus” strategy to reduce irrelevant interference and improve the translation of terminology.
Despite their effectiveness, these approaches inevitably require additional training, which incurs substantial computational cost. 
Moreover, fine-tuning often introduces the problem of catastrophic forgetting: a model adapted to rare-word corpora may experience degraded performance on general-domain data \cite{french1999catastrophic}.

\begin{figure*}[]
    \centering
    \includegraphics[width=1.0\linewidth]{struct_v4.png}
    \vspace{-4mm}
    \caption{
Illustration of the task vector ($\tau$) approach for addressing the problem in Section~\ref{sec:problem}. 
(a) A pre-trained model $\theta_0$ can be combined with a task vector $\tau_i$ to obtain a fused model $\hat{\theta}$ capable of recognizing the rare word $w_i$, without any fine-tuning. 
(b) Multiple task vectors can be flexibly combined to construct models that recognize multiple rare words. 
(c) Compared with a fine-tuned model $\theta_i$, a model built from $\tau_i$ not only captures the rare word recognition ability but also preserves the raw model's general capabilities, mitigating catastrophic forgetting.
}
    \label{fig:overview}
    \vspace{-4mm}
\end{figure*}

Limitations of fine-tuning arise not only in rare word tasks for speech models but also in other domains, such as large language models (LLMs).
Training LLMs such as Qwen \cite{yang2025qwen3} or Llama \cite{dubey2024llama} incurs substantial computational and financial costs. 
To enable a single large language model to acquire multiple capabilities without retraining, one promising direction is the task vector approach \cite{ilharco2022editing}. 
Task vectors represent the parameter shift induced by fine-tuning relative to a pretrained model. Arithmetic operations on these vectors enable the compositional transfer of knowledge across fine-tuned models.
Several extensions have been proposed to further improve this approach.
For example, the method in \cite{yadav2023ties} reduces interference and resolves conflicts in parameter signs by pruning and aligning parameter directions. Additionally, parameter pruning followed by the rescaling of remaining values helps to reduce redundancy and improve model merging \cite{yu2024language}.

Recently, the concept of task vectors has also been introduced into the speech domain. For example, the paper \cite{su2024task} proposes the use of \textit{SYN2REAL} task vectors, aiming to improve the robustness of ASR models trained on synthetic data when applied to real-world scenarios.  
Experiments in \cite{ramesh2024task,nagasawa2025task} demonstrate that task vectors derived from high-resource languages substantially improve performance on low-resource languages.
Meanwhile, several studies explore the application of task vectors in speech representation learning \cite{ritter2025distilling,lin2025speech}.
Given the diversity of rare words, repeatedly fine-tuning a new model for each rare word incurs prohibitive cost.
Therefore, it is worthwhile to investigate the effectiveness of leveraging task vectors to achieve training-free rare-word recognition and translation.

To address the challenges of high cost, catastrophic forgetting, and limited scalability in model fine-tuning for rare word speech recognition and translation, we propose a training-free approach based on task vectors. Our contributions are summarized as follows:
\begin{enumerate}
    \vspace{-1mm}
    \item To the best of our knowledge, this work is the first to apply task vector methods to speech-to-text models for rare word recognition. We address the challenges of high cost and catastrophic forgetting inherent in model fine-tuning.
    \vspace{-1mm}
    \item We introduce word-level task vector arithmetic, which enables direct composition of models trained on different rare words, thereby substantially enhancing scalability and reusability.
    \vspace{-2mm}
    \item We conduct extensive experiments across multiple rare-word domains, evaluating both recognition and translation tasks, providing empirical evidence of the effectiveness of our approach.
    \vspace{-2mm}
\end{enumerate}

\section{Method}
In this section, we begin by analyzing the limitations of fine-tuning in addressing rare-word recognition and translation.
We then present our proposed training-free method, which overcomes these challenges and provides a more scalable and flexible solution.

\subsection{Problem Formulation}
\label{sec:problem}
We study speech models for speech-to-text tasks, including ASR and AST. Let a pre-trained speech model be $\theta_0$. Given a set of rare word datasets $\{W_1, W_2, \dots, W_K\}$, where each subset $W_i$ contains $N_{W_i}$ samples involving only the rare word $w_i$, conventional fine-tuning requires: 
\begin{equation}
\vspace{-1mm}
\label{equ:1}
\theta_i = \underset{\theta}{\text{argmin}} \ \mathcal{L}(\theta_0; W_i), \quad \forall i \in [1,K],
\vspace{-1mm}
\end{equation}
where $\theta_i$ denotes the fine-tuned model derived from training $\theta_0$ on the subset $W_i$. This approach incurs a memory cost of $\mathcal{O}(1)$, and results in a training cost of $\mathcal{O}(N_{W_i})$ for each subset. This approach suffers from three fundamental limitations.

\textbf{High cost.} 
Enabling the raw model $\theta_0$ to recognize a rare word $w_i$ while preserving its general capabilities typically incurs a training cost much larger than $\mathcal{O}(N_{W_i})$. Moreover, as the number of rare words increases, the total cost grows linearly. In contrast, our goal is to incorporate new rare words with only $\mathcal{O}(1)$ additional training cost, without requiring any gradient updates. This process can be formally expressed as:
\begin{equation}
\vspace{-1mm}
\label{equ:2}
\hat{\theta} = \mathcal{F}(\theta_0, \tau_i) \quad \text{s.t.} \quad \frac{\partial\mathcal{F}}{\partial\theta} = 0.
\vspace{-1mm}
\end{equation}
In the above equation, $\tau_i$ represents the capability representation of the rare word $w_i$. By applying the function $\mathcal{F}$ to inject $\tau_i$ into the $\theta_0$, we obtain the fused model $\hat{\theta}$ with the ability to recognize $w_i$.

\textbf{Scalability constraint.}
When combining $K$ specialized models for different rare words, conventional ensemble methods incur a memory cost of $\mathcal{O}(K)$. Instead, we seek a fusion strategy \( \Phi \) that has scaling capabilities, such that
\begin{equation}
\vspace{-1mm}
\label{equ:3}
\hat{\theta} = \mathcal{F}(\theta_0, \Phi(\tau_1,\dots,\tau_K)),
\vspace{-1mm}
\end{equation}
where $\hat{\theta}$ acquires capabilities for handling multiple rare words. This approach reduces the memory cost to $\mathcal{O}(1)$.

\textbf{Catastrophic forgetting.}
Fine-tuning degrades general capabilities. Let $\mathcal{T}_{\text{gen}}$ be a generic task with loss $\mathcal{L}_{\text{gen}}$:
\begin{equation}
\vspace{-1mm}
\label{equ:4}
\Delta\mathcal{L}_{\text{gen}} = \mathcal{L}_{\text{gen}}(\theta_i) - \mathcal{L}_{\text{gen}}(\theta_0) \gg 0.
\vspace{-1mm}
\end{equation}
We require bounded degradation:
\begin{equation}
\vspace{-1mm}
\label{equ:5}
\left| \mathcal{L}_{\text{gen}}(\hat{\theta}) - \mathcal{L}_{\text{gen}}(\theta_0) \right| \leq \epsilon,
\vspace{-1mm}
\end{equation}
where $\epsilon$ quantifies the maximum allowable reduction in general task performance when the model gains the ability to recognize rare words.



\subsection{Training free method}
\label{sec:taskvector}


To address the above challenges, we introduce the task vector approach for rare words as Fig~\ref{fig:1}. 
We hypothesize that the recognition ability for rare words, obtained by fine-tuning a speech model on a rare word dataset, can be similarly encoded in task vectors. The task vector is defined as
\begin{equation}
\vspace{-1mm}
\label{equ:6}
\tau_i = \theta_i - \theta_0 ,
\vspace{-0mm}
\end{equation}
where $\tau_i$ captures the parameter differences between the fine-tuned and pretrained models, encoding the model’s recognition ability for the rare word $w_i$.

Building on this, we argue that integrating multiple rare word recognition capabilities can be achieved by strategically combining their respective task vectors. The overall integrated model is constructed as:
\begin{equation}
\vspace{-1mm}
\label{equ:7}
\hat{\theta} = \theta_0 + \Phi(\tau_1,\dots,\tau_K),
\end{equation}
where $\Phi(.)$ is a vector composition function designed to merge multiple task vectors without causing significant interference or performance loss. Considering the finer granularity of word-level task vectors, we explore and compare three distinct strategies:


\textbf{Task Arithmetic (TA) \cite{ilharco2022editing}.} A straightforward approach is vector addition. This method assumes that different task vectors can be combined linearly without significant conflict. The strategy is formally expressed as follows:
\begin{equation}
\vspace{-2.5mm}
\label{equ:8}
\Phi_{\text {Task Arithmetic}}(\tau_{1}, \tau_{2}, \ldots, \tau_{K})=\sum_{i=1}^{K} \alpha_{i} \tau_{i} ,
\end{equation}
where $\alpha_i$ controls the contribution of each vector.

\textbf{TrIm and Elect Sign (TIES) \cite{yadav2023ties}.}
This strategy is designed to address the problem of inconsistent parameter directions that can occur when combining multiple task vectors. The method includes three steps:
1) \textit{Trim}: We apply \cref{equ:trim} to sparsify each task vector. Specifically, for the parameters in $\tau_i$, we retain only the top-$p$ parameters with the largest magnitudes and set the remaining parameters to zero.
\begin{equation}
\vspace{-1mm}
\label{equ:trim}
\tau_i' = \mathrm{Trim}(\tau_i, p)
\vspace{-1mm}
\end{equation}
2) \textit{Elect Sign}: For each parameter index $j$, we determine the dominant direction by aggregating all trimmed task vectors as follows:
\begin{equation}
\vspace{-2mm}
\label{equ:electsign}
s^{*}[j] = \operatorname{sign}\!\Bigg(\sum_{i=1}^{K}\tau_i'[j]\Bigg).
\vspace{-0.5mm}
\end{equation}
3) \textit{Merge}: In the final fusion step, we aggregate only the parameter values that match the elected sign at each index \( j \), weighting them proportionally:
\begin{equation}
\vspace{-2mm}
\label{equ:11}
\Phi_{\text{TIES}}(\tau_1, \dots, \tau_K)[j] =
\sum_{i=1}^{K}\alpha_i \cdot \tau_i'[j] \cdot \mathbb{I}(\operatorname{sign}(\tau_i'[j]) = s^*[j]).
\vspace{-0.5mm}
\end{equation}


\textbf{Drop And Rescale (DARE) \cite{yu2024language}.} This strategy proceeds in two steps:
drop and rescale. For each task vector $\tau_i$, we independently drop each parameter with probability $p$:
\begin{equation}
\vspace{-2mm}
\tilde{\tau}_i = \mathrm{Drop}(\tau_i, p).
\vspace{-1mm}
\end{equation}
We then rescale the remaining parameters by a factor of $\frac{1}{1-p}$ to maintain consistent magnitude, and fuse them as
\begin{equation}
\vspace{-2mm}
\Phi_{\mathrm{DARE}}(\tau_1,\tau_2,\dots,\tau_K)
= \frac{1}{1-p} \sum_{i=1}^{K} \alpha_i \, \tilde{\tau}_i.
\vspace{-1mm}
\end{equation}

Overall, TIES and DARE can be viewed as advanced variants of Task Arithmetic, incorporating additional redundancy parameter pruning to reduce rare words interference.
\vspace{-1mm}
\section{EXPERIMENTAL SETUPS}
\label{sec:typestyle}

\subsection{Datasets}
We first used GPT-4o \cite{hurst2024gpt} to generate 10 rare words, including landmarks, locations, and individuals. These words were verified to be incorrectly recognized by the Whisper-medium model. For each rare word, GPT-4o generated over 1,000 contextualized English sentences, which were then translated into Chinese. Based on this bilingual corpus, we employed CosyVoice2 \cite{du2024cosyvoice} to synthesize speech, constructing a dataset with paired text and audio samples in the format \{en\_text, en\_speech, zh\_text, zh\_speech\}. For data partitioning, we randomly sampled 100 pairs per term as the test set, another 100 pairs as the validation set, and used the remaining samples for training. The detailed statistics of the rare word datasets are provided in Table~\ref{table:datasets}. In addition, we synthesized 1,702 pairs without rare words to create a general test set, which was used to evaluate the model’s generalization ability.


\begin{table}[ht]
\vspace{-5mm}
\centering
\caption{Statistics of the rare word datasets.}
\label{table:datasets}
\begin{tabular}{clcc}
\toprule
\textbf{Domain} & \textbf{Rare Word} & \textbf{Train} & \textbf{ID}\\
\midrule
\multirow{3}{*}{Landmark} & Neuschwanstein Castle & 1160 &$W_1$\\
                           & Berlin Wall           & 1345&$W_2$ \\
                           & Milan Cathedral       & 1359&$W_3$ \\
\midrule
\multirow{4}{*}{Location} & Triberg Waterfall     & 1248&$W_4$ \\
                           & The Rhine Valley      & 1500&$W_5$ \\
                           & Avignon               & 1414&$W_6$ \\
                           & Carcassonne           & 1363&$W_7$ \\
\midrule
\multirow{3}{*}{Individual} & Raphael             & 1262&$W_8$ \\
                             & Goethe              & 1148&$W_9$ \\
                             & Voltaire            & 1372&$W_{10}$ \\
\bottomrule
\end{tabular}
\vspace{-2mm}
\end{table}

\subsection{Training configuration}

All experiments are conducted on the Whisper-medium \cite{radford2023robust} model. Whisper is a multitasking system capable of performing both ASR and AST. In the AST setting, it directly translates speech from multiple source languages into English text.
For all fine-tuning strategies considered in this work, we train for 30 epochs with an initial learning rate of $1 \times 10^{-3}$. We select the checkpoint with the best validation performance for evaluation and downstream task vector construction. For \textbf{Task Arithmetic}, \textbf{TIES}, and \textbf{DARE}, we normalize the hyperparameters such that $\sum_{i=1}^{K} \alpha_i = 1$ with $\alpha_1 = \alpha_2 = \cdots = \alpha_K$.
For both \textbf{TIES} and \textbf{DARE}, the pruning probability $p$ is fixed at $0.5$.

We consider three types of baseline models:

\begin{enumerate}
    \vspace{-1.3mm}
    \item \textbf{Raw}: directly evaluating the publicly available pretrained checkpoint without any additional fine-tuning;
    \vspace{-1.3mm}
    \item \textbf{Fine-tuned (FT)}: further fine-tuning the pretrained model on the rare word dataset before evaluation;
    \vspace{-1.3mm}
    \item \textbf{Average}: constructing a new model by directly averaging the parameters of multiple models, serving as a reference for comparison with the three composition strategies introduced in Section~\ref{sec:taskvector}.
    \vspace{-1.3mm}
\end{enumerate}

We conduct systematic experiments to validate the effectiveness of task vectors. The setup is as follows:

\textbf{(1) Single task vector.}  
We start by assessing the basic effectiveness of task vectors. Specifically, we fine-tune 10 models on 10 different rare word datasets for Chinese-to-English speech translation. These fine-tuned models are then combined using three fusion strategies introduced in Section~\ref{sec:taskvector} to create task-vector-based rare word models. The Average strategy simply takes the average of the parameters from the fine-tuned model and the raw model. It is important to note that in this setting, both the fine-tuned models and the fusion models are restricted to handling one single rare word at a time.

\textbf{(2) Composition of multiple task vectors.}
We investigate whether combining multiple task vectors enables simultaneous recognition of several rare words in Chinese-to-English speech translation.
Under task vector strategies or the Average method, “1 rare word” uses a single fine-tuned model, “2 rare words” fuses two fine-tuned models, and so on. 
By contrast, the fine-tuning baseline jointly trains a single model on all selected rare word datasets.

\textbf{(3) Mitigating catastrophic forgetting.}  
Fine-tuning a general-purpose speech model on specific domains often degrades its basic capabilities. To study this, we further evaluate the models from \textbf{Single task vector} on their corresponding rare word test sets, measuring performance on Chinese speech transcription. 
\vspace{-2mm}
\subsection{Evaluation}

For translation performance, we adopt SacreBLEU \cite{post2018call} as the primary evaluation metric. To ensure consistency, we normalize both the model outputs and reference texts to lowercase before calculating the BLEU scores.
For transcription performance, we use Character Error Rate (CER) as the evaluation metric for Chinese speech transcription, which measures accuracy at the character level.
\vspace{-2mm}
\section{RESULTS AND DISCUSSION}
\label{sec:results}

\subsection{Single task vector}
\label{section:5.1}

The results of single task vector are summarized in Table~\ref{table:1}.
Overall, TA, TIES, and DARE perform on par with fine-tuned models on most rare words and even surpass them in certain cases (e.g., “Neuschwanstein Castle ($W_1$)” and “Milan Cathedral ($W_3$)”). On the general dataset, these strategies improve performance by about 5 BLEU over the Raw model and 2 BLEU over fine-tuned models, demonstrating clear advantages.
Fine-tuned models often introduce parameter redundancy. Adjusting task vector coefficients or increasing drop probability can enhance generalization on the general dataset without sacrificing rare word performance.
Additionally, the Average method performs nearly identically to TA, as it can be viewed as its special case with equal weights of $\alpha_i$, while TA allows flexible coefficient adjustment.

\begin{table}[t]
\centering
\caption{BLEU ($\uparrow $) scores on the Chinese-to-English speech translation task for 10 rare word datasets. For each word, the best performance among ``Average'', ``TA'', ``TIES'', and ``DARE'' is highlighted in \textbf{bold}. Results on the general-domain test set are also reported.}
\label{table:1}
\begin{tabular}{
    l@{\hspace{2pt}}  
    c@{\hspace{2pt}}
    c
    c
    c
    c
    c
}
\toprule
\textbf{Words} & \textbf{Raw} & \textbf{FT} & \textbf{Average} & \textbf{TA} & \textbf{TIES} & \textbf{DARE} \\
\midrule
$W_1$ & 25.28 & 66.20 & 66.57 & \textbf{66.77} & 63.76 & 66.57 \\
$W_2$ & 35.79 & 74.65 & \textbf{74.88} & \textbf{74.88} & 74.25 & \textbf{74.88} \\
$W_3$ & 42.13 & 67.81 & \textbf{68.66} & \textbf{68.66} & 66.05 & 68.63 \\
$W_4$ & 44.49 & 71.73 & \textbf{73.17} & \textbf{73.17} & 72.44 & 73.07 \\
$W_5$ & 24.29 & 63.30 & 59.46 & 59.46 & \textbf{60.01} & 59.46 \\
$W_6$ & 22.56 & 70.83 & \textbf{70.66} & \textbf{70.66} & 68.44 & 70.00 \\
$W_7$ & 28.32 & 59.36 & \textbf{61.06} & 61.02 & 59.21 & 60.53 \\
$W_8$ & 28.46 & 63.48 & 62.53 & 62.60 & 59.33 & \textbf{62.86} \\
$W_9$ & 38.08 & 70.11 & \textbf{68.34} & \textbf{68.34} & 68.21 & 67.86 \\
$W_{10}$ & 32.91 & 71.19 & 64.82 & 64.82 & 63.43 & \textbf{64.96} \\
\midrule
General & 35.66 & 37.61 & 40.64 & 40.76 & \textbf{40.77} & 40.76 \\
\bottomrule
\end{tabular}
\end{table}

\subsection{Multiple task vectors}
\label{section:5.2}
\begin{figure}[t]
    \vspace{-2mm}
    \centering
    \includegraphics[width=0.92\linewidth]{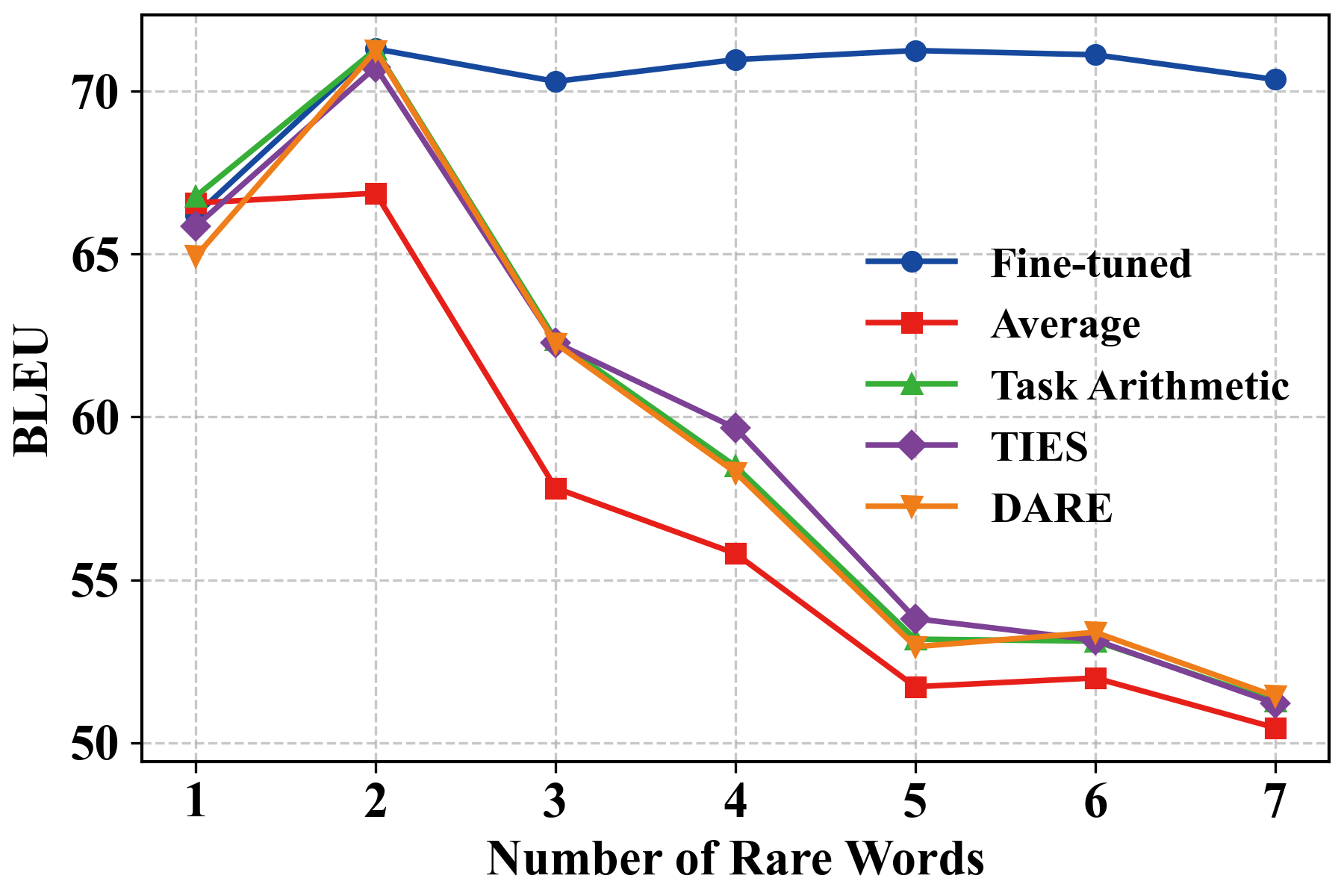}
    \caption{BLEU scores on the Chinese-to-English speech translation task as the number of rare words increases.}
    \label{fig:1}
    \vspace{-6mm}
\end{figure}

This subsection investigates how the performance of task-vector-based models evolves as the number of rare word task vectors increases. The experiment results are shown in Figure~\ref{fig:1}. The results indicate that when the number of rare words is 1, all fusion strategies achieve performance comparable to that of the corresponding fine-tuned model. Overall, as the number of rare words increases, the performance of the Average, TA, TIES, and DARE declines. In particular, within the range of 2 to 4 rare words, the three task-vector–based strategies consistently outperform the Average baseline. However, this advantage diminishes as the number of rare words continues to rise.
We attribute this decline to the distributional differences across fine-tuned models. Simple parameter averaging tends to introduce interference between tasks. In contrast, strategies such as TA, which adjust task weights via coefficients $\alpha_i$, or TIES and DARE employ drop probabilities $p$ to suppress redundant parameters, are more effective in mitigating cross-task conflicts. Nonetheless, as the number of task vectors further increases, the likelihood of parameter conflicts inevitably rises, which explains the gradual weakening of these strategies.

\subsection{Catastrophic forgetting}
\label{section:5.3}

\begin{table}[t]
\centering
\caption{CER (\%) ($\downarrow$) scores on the Chinese speech recognition task for the 10 rare word datasets. The best CER for each word is highlighted in \textbf{bold}, selected only among the four fusion strategies.}
\label{table:2}
\begin{tabular}{l
                r
                r
                r
                r
                r
}
\toprule
\textbf{Words} & \textbf{Raw}  & \textbf{Average} & \textbf{TA} & \textbf{TIES} & \textbf{DARE} \\
\midrule
$W_1$ & 10.00 & 19.74 & 6.60 & \textbf{5.92} & 6.60 \\
$W_2$ & 10.06  & 29.23 & 29.23 &\textbf{21.75} & 29.42 \\
$W_3$ & 7.34  & 14.56 & 6.23 & \textbf{4.96} & 5.79 \\
$W_4$ & 6.72  & 24.33 & 16.76 & \textbf{6.50} & 10.63 \\
$W_5$ & 9.03  & 17.47 & 17.47 & \textbf{9.93} & 17.61 \\
$W_6$ & 11.26  & \textbf{4.45} & \textbf{4.45} & 4.88 & \textbf{4.45} \\
$W_7$ & 9.17  & 21.48 & 7.42 & \textbf{7.06} & 7.62 \\
$W_8$ & 8.68  & 14.18 & 7.55 & \textbf{5.96} & 8.22 \\
$W_9$ & 12.78  & 25.72 & 25.72 & \textbf{16.10} & 25.72 \\
$W_{10}$ & 11.60  & 16.64 & 16.64 & \textbf{13.74} & 15.88 \\
\bottomrule
\end{tabular}
\vspace{-4mm}
\end{table}

A common challenge with fine-tuning large speech models on domain-specific tasks is the degradation of their general capabilities. 
Building on the models in Table~\ref{table:1}, we evaluate their Chinese transcription performance on the rare word test sets (Table~\ref{table:2}). From these results, the original Whisper model shows strong multi-task ability, handling both translation and recognition. However, fine-tuning for rare word translation, while boosting target performance, causes catastrophic forgetting—Chinese speech recognition collapses entirely. 
In contrast, task-vector-based strategies mitigate this degradation to varying degrees. Among them, TIES achieves the lowest CER on most test sets, with especially strong results for “Berlin Wall ($W_2$)” and “Raphael ($W_8$)”. In “Carcassonne ($W_7$)”, it even outperforms the raw model, suggesting that it not only preserves but sometimes improves recognition.
This success stems from TIES’s precise modeling of task-specific parameter updates. Its task vector $\tau_i$ captures the translation ability for the rare word $w_i$ and, when combined with the raw model, strengthens translation while preserving recognition. In contrast, Average, TA, and DARE show less stable parameter fusion, resulting in higher CER.

Overall, task-vector-based parameter fusion effectively mitigates catastrophic forgetting, enabling models to achieve strong task performance while maintaining general abilities.


\section{CONCLUSION}
\label{sec:conclusion}


In this work, we addressed the challenges of rare word recognition and translation in speech-to-text systems, where direct fine-tuning suffers from high cost, catastrophic forgetting, and poor scalability. We proposed a training-free paradigm based on task vectors, and introduced word-level task vector arithmetic to flexibly compose rare-word capabilities. Experiments across multiple domains demonstrated that our approach not only achieves competitive or superior performance on target words compared to fine-tuned models, but also improves general test performance while alleviating forgetting. These results highlight task vectors as a practical and scalable solution for handling rare entities such as place names, personal names, and technical terms in real-world applications.




\vfill\pagebreak

\bibliographystyle{IEEEbib}
\bibliography{strings,refs}

\end{document}